# Unprecedented superionicity of ultra-low barrier in $A_{0.5}CoO_2$ (A=Li, Zn)

Xuechen Wang[1], Yaxin Gao[2], Menghao Wu[1*]

[1] School of Physics and School of Chemistry, Huazhong University of Science and Technology, Wuhan, Hubei 430074, China

[2] School of Physics and Mechanical Electrical & Engineering, Hubei University of Education, Wuhan, Hubei 430205, China.

*Email: wmh1987@hust.edu.cn

Abstract   The ion conductivity of a solid-state ion conductor generally increases exponentially upon reduction in ion migration barrier. For prevalent cathode material $LiCoO_2$, the room-temperature ion conductivity and migration barrier are respectively around $10^{-4}$ S/cm and 0.3 eV. In this paper, through first-principles calculations we predict the existence of 1D superionicity as the Li ions in O2 $LiCoO_2$ are transformed to $Zn_{0.5}CoO_2$ or $Li_{0.5}CoO_2$ via cation-exchange reaction or deintercalation. The ion migration barriers (0.01-0.02 eV) even lower than room-temperature $\sim k_BT$ are reduced by more than an order of magnitude compared with $LiCoO_2$, which are facilitated by facile transition of mobile ions between two coordination configurations. The room-temperature ion conductivity is estimated to be over 50 S/cm, enhanced by 2-3 orders of magnitude compared with current highest reported value. Such unprecedented superionicity may also exist in other similar layered ion conductors, which may render technical advances and exotic effects such as ultrafast ion batteries and quantized ferroelectricity.

Solid-state ion conductors exhibiting fast ion transport have been widely used in energy-storage devices like batteries.[1-8] In particular, lithium-ion batteries have emerged as an essential component of modern portable power sources. $LiCoO_2$ is one of the most extensively studied cathode materials due to its high energy density and long cyclability,[9] where the layered structure is formed by $Co^{3+}$-based edge-sharing $CoO_6$ octahedrons separated by lithium layers, facilitating the intercalation/deintercalation of Li ions. However, its low room-temperature ion conductivity (~$10^{-4}$ S/cm) [10] due to high migration barrier (~0.3-0.4 eV) [11-13] would cause not only a sluggish transport of Li ions, but also a large polarization at the cathode-electrolyte interface. The ion conductivity is proportional to $\exp\left(-\frac{\Delta}{k_B T}\right)$, which can be greatly enhanced even when the barrier $\Delta$ is only slightly reduced. For example, the room-temperature ion conductivity and migration barrier for $LiNbOCl_4$ have been respectively estimated to be 10.4 mS/cm and 0.205 eV[14], which are 24 mS/cm and 0.155 eV for $Li_{5.3}PS_{4.3}ClBr_{0.7}$ with disorder between $Cl^-$, $Br^-$ and $S^{2-}$ sites[15]. In 2023, Ryoji Kanno et al. synthesized $Li_{9.54}[Si_{1-\delta}M_\delta]_{1.74}P_{1.44}S_{11.1}Br_{0.3}O_{0.6}$ (M = Ge, Sn) with a high room-temperature conductivity of 32 mS/cm and low migration barrier of 0.09 eV, surpassing all previous records for inorganic Li-ion conductors[16]. However, even such a low barrier still corresponds to $k_B T$ of a high temperature over 1000K, surpassing the melting points of many solid-state ion conductors. If the barrier can be reduced below ~$k_B T$, ion can move as quasi-free particle and the potential wells provide scattering instead of localization, which may give rise to various exotic effects described in lattice gas model.[17]

Additionally, some solid-state ion conductors with ion vacancies may exhibit unconventional ferroelectricity with quantized polarizations.[18] It has also been proposed that such quantized ferroelectricity may even emerge in monovalent ion conductors with non-ferroelectric crystal lattices if their mobile monovalent cations are substituted by multivalent cations (e.g., $Mg^{2+}$, $Al^{3+}$)[19] via cation-exchange reaction, based on previous experimental reports on topotactic reaction with an aliovalent cation between trigonal layers of ion conductors.[20, 21] However, the experimental realization of such quantized ferroelectricity is also hindered by the high migration barriers of those ion conductors (e.g.,

0.31 eV per Na ion for $Na_4SnS_4$,[18] and 0.29 eV per Mg ion for $Mg_{0.5}CrS_2$,[19] much higher than the switching barrier of $BaTiO_3$ around 0.1 eV). In this paper, we predict by *ab initio* calculations that when the Li ions in O2 $LiCoO_2$ are substituted by half number of divalent $Zn^{2+}$ by cation-exchange reaction, or half of them are deintercalated, arrays of 1D ion conduction channels with ultra-low migration barriers (0.01-0.02 eV) will be formed, giving rise to unprecedented room-temperature ion conductivity over 50 S/cm. The ultralow migration barrier can be attributed to the existence of two interconvertible states with different coordinate configurations that are almost degenerate in energy, making the intercalated ions behave like 1D liquids at ambient conditions.

Our theoretical calculations are performed based on density-functional-theory (DFT) methods implemented in the Vienna Ab initio Simulation Package (VASP) code[22, 23]. The generalized gradient approximation (GGA) in the form of Perdew-Burke-Ernzerhof functional (PBE)[24] and the projector augmented wave (PAW)[25] method are applied to describe the exchange-correlation and the core electrons, respectively. PBE+ U method with an effective U value ($U_{eff}$) of 4.9 eV[26] is used to account for strong correlation. The kinetic energy cutoff was set at 500 eV, and the Brillouin zone was sampled by 7 × 5 × 3 K points for O2-state and 5 × 7 × 5 K points for O3-state using the Monkhorst−Pack scheme[27]. The convergence threshold for self-consistent-field iteration is set to $10^{-6}$ eV and all structures are fully optimized until the forces are smaller than 0.01 eV/ Å. The Berry phase method was employed to evaluate crystalline polarizations, and the ion migration pathways were obtained by using the solid-state nudged elastic band (SSNEB)[28] method.

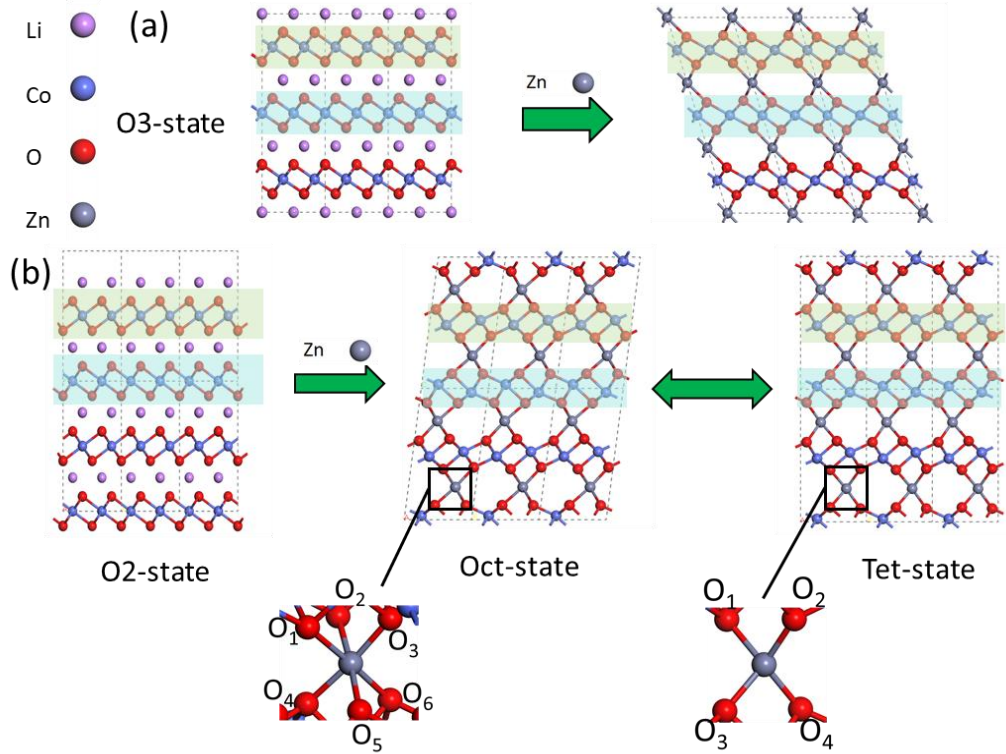

Figure 1. (a) Transformation from O3 state LiCoO$_2$ to Zn$_{0.5}$CoO$_2$ upon cation-exchange reaction. (b) Transformation from O2 state LiCoO$_2$ to Oct and Tet state Zn$_{0.5}$CoO$_2$, respectively with distorted ZnO$_6$ octahedrons and ZnO$_4$ tetrahedrons.

The experimentally verified O3 and O2 phase LiCoO$_2$ respectively with parallel and antiparallel CoO$_2$ layers are displayed in Fig. 1, where the Li ions are intercalated into the interstitial octahedral sites.[29-32] As the Li ions are substituted by half number of Zn ions, the CoO$_2$ layers in O2 phase Zn$_{0.5}$CoO$_2$ become buckled in the ground state (see the comparison in energy between different configurations in Fig. S1), which are still planar in O3 phase. Two configurations may form in O2 Zn$_{0.5}$CoO$_2$, where each Zn ion can either bond with 6 O atoms in a ZnO$_6$ octahedron (denoted as Oct-state), or 4 O atoms in a ZnO$_4$ tetrahedron (denoted as Tet-state, only 2 meV/f.u. higher in energy compared with Oct-state). The bonding configurations of Zn ions in two phases are shown in Fig. 1(b), where the Zn-O bond lengths in Tet-state range from 1.998 to 2.018 Å ($d_{Zn-O1}$= $d_{Zn-O3}$=1.998 Å, $d_{Zn-O2}$= $d_{Zn-O4}$=2.018 Å), mostly shorter compared with Oct-state ($d_{Zn-O1}$= $d_{Zn-O2}$ =2.226 Å, $d_{Zn-O5}$= $d_{Zn-O6}$=2.195 Å, $d_{Zn-O3}$=2.015 Å, $d_{Zn-O4}$=2.043 Å), and the differences between those bond lengths reveal the distortions of octahedrons and tetrahedrons.

As shown in Fig. 2 where adjacent in-plane O atoms of each layer are linked, each Zn cation in LiCoO$_2$ is located between two antiparallel oxygen triangles of adjacent layers respectively marked in red and blue, with overlapping centers from the overview. As Li ions are substituted by half number of Zn ions, the deviations between the centers of two triangles upon slightly interlayer sliding make them staggered in Oct phase Zn$_{0.5}$CoO$_2$, accompanied by the distortions of ZnO$_6$ octahedrons. According to our NEB calculations, the transition barrier between Oct and Tet-phase upon displacements of Zn ions (accompanied by slightly interlayer sliding) is only around 20 meV/f.u..

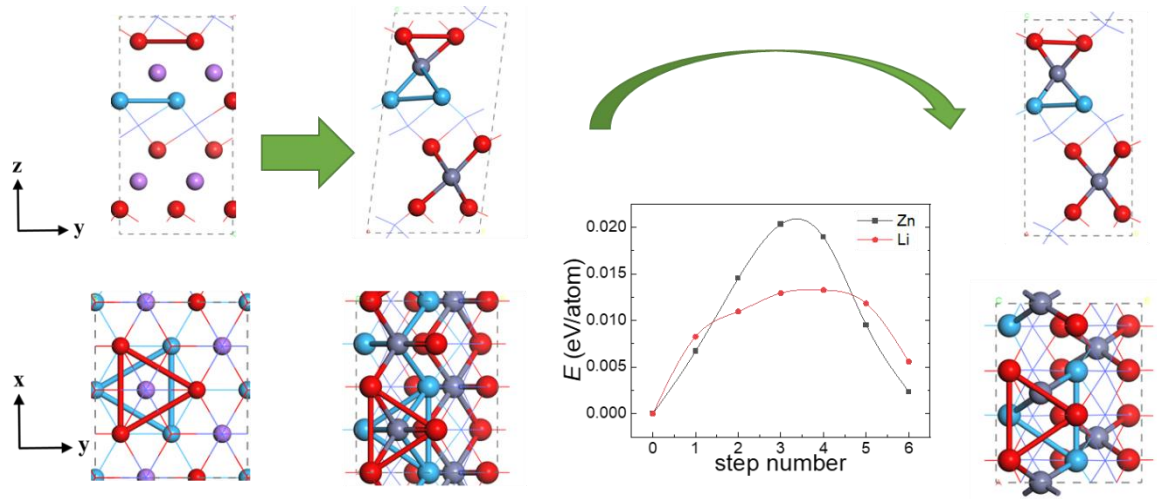

Figure 2. The side view and overview of O2 LiCoO$_2$, Oct and Tet state of ZnCo$_2$O$_4$, and Oct-Tet phase transition pathways for ZnCo$_2$O$_4$ and LiCo$_2$O$_4$, where the upper and lower oxygen triangles are marked in red and blue, respectively.

It is noteworthy that when half of Li ions in O2 LiCoO$_2$ are deintercalated, similar Oct and Tet-phase for Li$_{0.5}$CoO$_2$ may emerge with a small energy difference around 5 meV/f.u. between them, and their transition barrier of 13 meV/f.u. is even lower, although it is metallic compared with insulating Zn$_{0.5}$CoO$_2$ with a bandgap of 1.27 eV. Similar low migration barrier may also exist in some other layered ion conductors like Zn$_{0.5}$NiO$_2$ (see the estimated value around 33 meV in Fig. S2). Compared with the ground state with pyrochlore lattices, such layered structures are respectively 64 and 36 meV/atom for Zn$_{0.5}$CoO$_2$ and Zn$_{0.5}$CoO$_2$, so their stable formations via cation-exchange reaction from O2 LiCoO$_2$ should be still possible. According to our *ab initio* molecular dynamic (AIMD) simulations in the canonical ensemble with the temperature being controlled at 600 K in Fig. S3, the robust integrity of layered structure suggests good thermal stability.

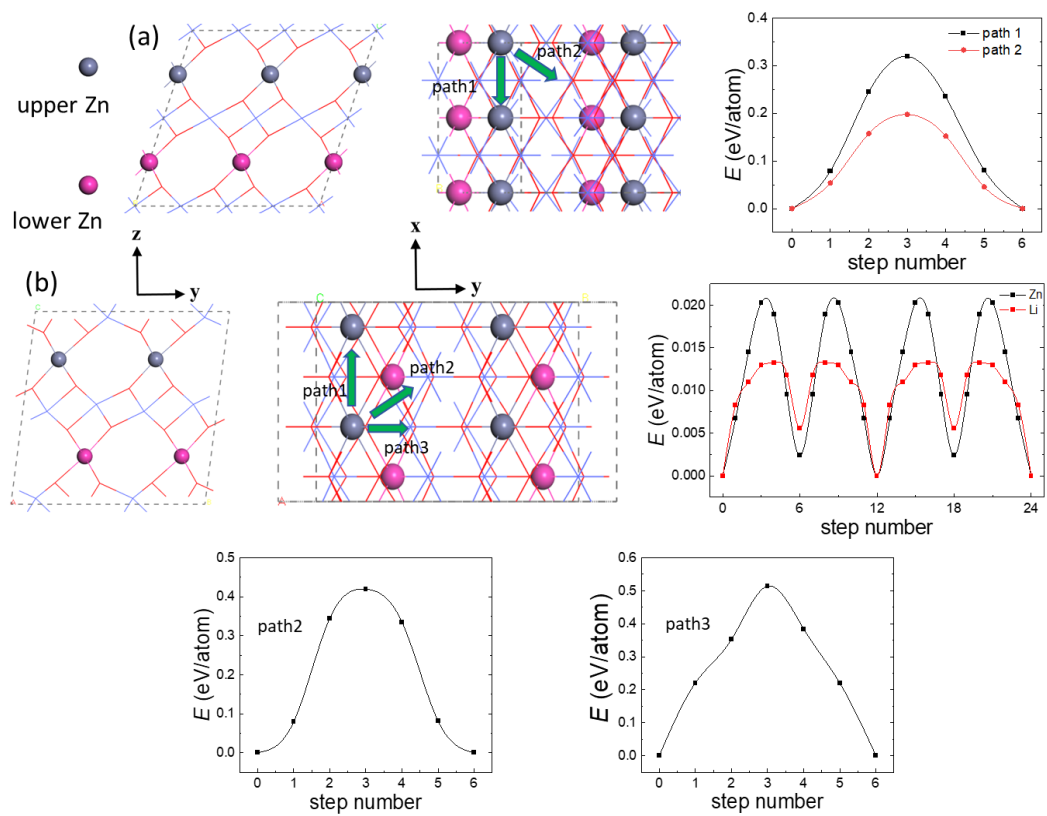

Figure 3. (a) The sideview and overview structure of O3-ZnCo$_2$O$_4$ and two ion migration pathways. (b) The sideview and overview structure of O2-ZnCo$_2$O$_4$ and three ion migration pathways, where the migration along path1 also gives ultralow barrier for O2-LiCo$_2$O$_4$.

The migration barrier of Li ions in O3 LiCoO$_2$ are typically ~0.3-0.4 eV according to previous studies, [11, 12] which is reduced to ~0.2-0.3eV per Zn ion upon cation-exchange reaction, as shown in Fig. 3(a). In the anti-parallel stacking configuration, the migration of Zn ions along –x direction can be greatly facilitated by the low barrier of Oct-Tet phase transition where the Zn ions are displaced by half lattice constant |**a**|/2, and the migration for one lattice constant |**a**| is equivalent to Oct-Tet-Oct phase transition. As shown in Fig. 3(b), the migration barrier along the conduction channel is just the Oct-Tet phase transition barrier, even lower than ~$k_BT$ at ambient conditions and more than an order of magnitude lower compared with migration barrier in O3 phase. Meanwhile the migration barriers along other two directions (marked as path 2 and 3 in Fig. 3(b)) are much higher (all over 0.4 eV). Such anisotropy should clarify the formations of confined 1D ion liquids in Fig. S3, where the displacements of Zn ions along –y direction at elevated temperature are negligible compared with –x direction.

A coarse estimation of ion conductivity ($\delta_{ion}$)[33] can be formulated as

$$\delta_{ion} = \left(\frac{\gamma c \nu Z^2 e^2 a_0^2}{k_B T}\right) \exp\left(-\frac{\Delta G_m}{k_B T}\right)$$

where Z = 2 is the charge number of Zn ion, $a_0$ = 3.03×10$^{-8}$ cm and $\Delta G_m$ = 0.02 eV are respectively the transfer distance and migration barrier of each Zn ion. c = 0.012×10$^{24}$ cm$^{-3}$ is its ion density, $\nu$ = 10$^{13}$ s$^{-1}$ is the frequency and $\gamma$ = 1 represents the 1D Zn ion conduction channel. Given the parameters above, $\delta_{ion}$ is estimated to be 55.6 S/cm for ZnCo$_2$O$_4$. Similarly, the Li ion conductivity in O2 LiCo$_2$O$_4$ estimated to be 27.3 S/cm is not only 5 orders of magnitude larger than $\delta$ in O3 LiCoO$_{2[10]}$ (10$^{-4}$ S/cm), but also much higher than all superionic crystals currently known, like lithium argyrodites[34] (~10 mS/cm), α-RbAg$_4$I$_5$[35] (0.3 S/cm) and α-KAg$_4$I$_5$ (0.221 S/cm)[36]. However, such ultrahigh conductivity is highly anisotropic since the migration barrier is much higher along two other directions, where the coordination phase transition is not involved.

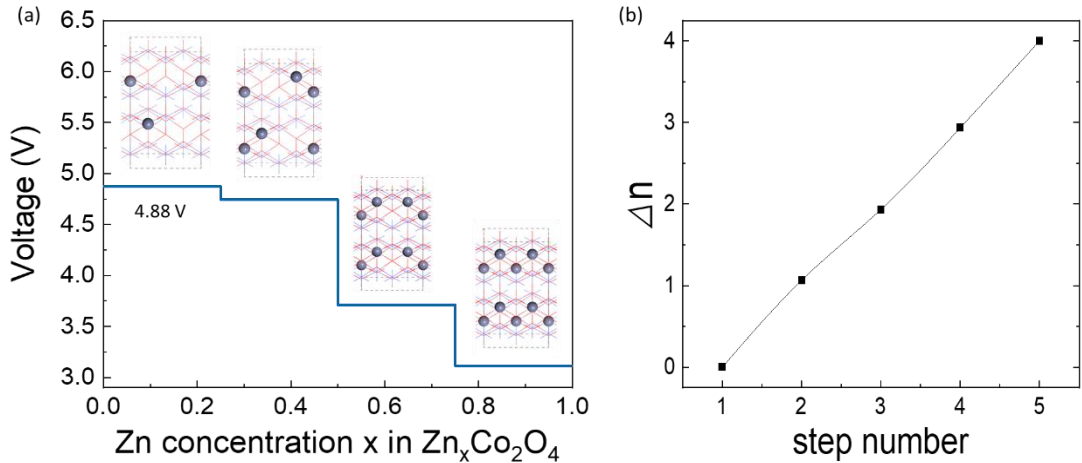

Figure 4. (a) Calculated open-circuit voltage (OCV) as a function of Zn-ion concentration x in Zn$_x$Co$_2$O$_4$. (b) The polarization quantum evolution upon the migration of Zn ions along –x direction for one lattice constant |**a**|, where $\Delta n = \Omega \Delta P / e|\mathbf{a}|$.

We use the following equation for a rough estimation of the specific capacity of ZnCo$_2$O$_4$

as a cathode material:

$$C = \frac{nZF}{3.6m}$$

where n = 1 is the mole number of Zn ion, Z = 2 is the charge number of Zn ion, F is the Faraday constant (F = 96485 C·mol$^{-1}$ ), and m = 247.24 g·mol$^{-1}$ is the mole mass of ZnCo$_2$O$_4$. Given the parameters above, C is estimated to be 216.44 mAh/g. The open-circuit voltage (OCV) of ZnCo$_2$O$_4$ in Fig. 4(a) estimated to be 4.88V is larger than most Li-ion cathode materials such as LiCoO$_2$ (3.75 V) and LiFePO$_4$ (3.47 V)[11], and the estimated energy density of 1056.3 Wh/kg is also higher (LiCoO$_2$:760 /1027.5 Wh/kg in experimental/theoretical reports; LiFePO$_4$: 476 /590 Wh/kg in experimental/theoretical reports) [37]. With high energy density and ultra-low migration barrier, ZnCo$_2$O$_4$ could be a promising material for ultra-fast rechargeable ion battery.

Finally, the low ion migration barrier of ZnCo$_2$O$_4$ should facilitate the realization of unconventional ferroelectricity with quantized polarization, considering the unattainable voltage required for switching due to the high barriers over 0.3 eV in previous predicted systems.[18, 19] Upon migration of Zn ions for one lattice constant along -x direction corresponding to 1 ion vacancy at each channel, the change of polarization quanta (Δ$n$=4 in Fig. 4(b)) is 63.6 μC/cm$^2$, already two times larger compared with BaTiO$_3$. In a ZnCo$_2$O$_4$ crystal with size ~1 μm and a density of only 1 vacancy per 100 Zn ions, the migration distance in each channel will be orders of magnitude larger, giving rise to an immense ferroelectric polarization around ~1000 μC/cm$^2$, with barriers comparable to the lowest values of current ferroelectrics.[38, 39]

In summary, we predict the existence of superionicity in antiparallel stacking ZnCo$_2$O$_4$ and LiCo$_2$O$_4$ based on O2 LiCoO$_2$. The ultralow transition barriers (0.01-0.02 eV) between Oct and Tet states are also the ion migration barrier along the 1D conduction channels, leading to the highest solid-state ion conductivity known to date. Similar superionicity may also exist in some other layered ion conductors like ZnNi$_2$O$_4$. Considering the technical advances and various exotic effects[17] that may be brought by such elusive superionicity with barrier lower than room-temperature ~$k_B$T, our prediction should deserve further experimental efforts.

Acknowledgement

This work is supported by National Natural Science Foundation of China (Nos. 22073034).

Supporting Information